\documentclass[twocolumn]{aastex62}

\accepted{02 Jan 2021 in The Astrophysical Journal Letters}

\begin{document}

\title{The Hubble WFC3 Emission Spectrum of the Extremely-Hot Jupiter, KELT-9\,b}

\author[0000-0001-6516-4493]{Quentin Changeat$^*$}
\affil{Department of Physics and Astronomy, University College London, London, United Kingdom}
\altaffiliation{These authors contributed equally to this work.}
\author[0000-0002-5494-3237]{Billy Edwards$^*$}
\affil{Department of Physics and Astronomy, University College London, London, United Kingdom}
\thanks{These authors contributed equally to this work.}
\altaffiliation{These authors contributed equally to this work.}

\begin{abstract}
    Recent studies of ultra hot-Jupiters suggested that their atmospheres could have thermal inversions due to the presence of optical absorbers such as TiO, VO, FeH and other metal hydride/oxides. However, it is expected that these molecules would thermally dissociate at extremely high temperatures, thus leading to featureless spectra in the infrared. KELT-9\,b, the hottest exoplanet discovered so far, is thought to belong to this regime and host an atmosphere dominated by neutral Hydrogen from dissociation and atomic/ionic species. Here, we analysed the eclipse spectrum obtained using the Hubble Space Telescope's Wide Field Camera 3 (WFC3) and, by utilising the atmospheric retrieval code TauREx3, found that the spectrum is consistent with the presence of molecular species and is poorly fitted by a simple blackbody. In particular, we find that a combination of TiO, VO, FeH and H- provides the best-fit when considering HST, Spitzer and TESS datasets together. Aware of potential biases when combining instruments, we also analysed the HST spectrum alone and found that TiO and VO only were needed in this case. These findings paint a more complex picture of the atmospheres of ultra-hot planets than previously thought. \vspace{15mm}
\end{abstract}

\section{Introduction}

Models suggest that, in the atmospheres of ultra-hot Jupiters, molecular compounds should evaporate. Some of these, such as titanium oxide (TiO) and vanadium oxide (VO), were expected to efficiently absorb incident flux high in the atmospheres of these planets, instigating an increase in temperature \citep[e.g.][]{Hubeny_thermal_inv,Fortney_2008}. Studies have since searched for these thermal inversions, initially by comparing eclipse depths between the Spitzer IRAC 3.6 and 4.5 $\mu$m channels and more recently by Hubble's WFC3 \citep[e.g.][]{arcangeli_wasp18_em,mansfield_hatp7,Evans_wasp121_e2,Edwards_2020_ares,pluriel_aresIII}. 

KELT-9\,b orbits an A0V/B9V star (T = 10140 K) and, with a day-side temperature of $\sim$4500 K, is itself hotter than many stars \citep{gaudi_k9}. Given the extreme temperatures, the majority of molecules are anticipated to be dissociated leaving only atomic species \citep{kitzmann_k9}. Ground-based high resolution observations have detected a number of metals including iron, titanium and calcium \citep{hoeij_k9,hoeij_k9_2,yan_k9, cauley_k9, turner_k9, pino_k9}. Observations of KELT-9\,b with TESS and Spitzer have revealed an asymmetric transit \citep{ahlers_k9}, induced by the fast rotation of its host star. The rotation leads to a non-uniform structure in the star, which has a larger and brighter equator than the poles, whereas KELT-9\,b orbits with a 87$^{\circ}$ spin-orbit angle. The planet has a low day-night temperature contrast with indications for H$_2$ dissociation and recombination \citep{wong_k9,mansfield_k9}. It is subject to intense irradiation from the star and extreme atmospheric escape due to its large extended hydrogen envelope reaching the Roche-Lobe limit \citep{Yan_2018, Wyttenbach_2020}. Measurements of the neutral iron line by \citet{pino_k9} imply the presence of a thermal inversion, as does modelling by \citet{fossati_2020_datadriven}.

Given the temperature of KELT-9\,b, its day-side emission spectrum was anticipated to resemble a blackbody \citep{wong_k9,Lothringer_2018}. Here we analyse the Hubble WFC3 emission spectrum of KELT-9\,b and find that, contrary to these predictions, the day-side spectrum deviates significantly from a blackbody. We perform atmospheric retrievals to attempt to explain these features, finding optical absorbers are required, and we compare our results to the abundances expected from equilibrium models.

\section{Hubble Data Reduction and Analysis}

The Hubble WFC3 eclipse observation of KELT-9\,b was acquired as part of proposal 15820 (PI: Lorenzo Pino, \citet{Lorenzo_2019_proposals}). Hubble attempted to acquire the data on 21$^{\rm st}$ November 2019 but there were issues with the guidance, with a failure in guide star acquisition and the pointing was completed using gyros only, making the data unusable. The observation was successfully repeated on 25$^{\rm th}$ January 2020.

The visit utilised the GRISM512 aperture and SQ512 subarray, with an exposure time of 92.538 s which consisted of 6 up-the-ramp reads using the SPARS25 sequence. The visit had a scan rate of 0.438 "/s, resulting in a scan length of 43.559 " which stretches over approximately 335 pixels, one of the longest spatial scans completed thus far for exoplanet spectroscopy after 55\,Cancri\,e (350 pixels, \citet{tsiaras_55cnce}). Both forward and reverse scans were used to increase the duty cycle. Additionally, a single F164N direct image was taken at the being of each orbit for wavelength calibration.

We reduced the data using the Iraclis\footnote{\url{https://github.com/ucl-exoplanets/Iraclis}}, open-source software for the analysis of WFC3 scanning observations \citep{tsiaras_hd209} and the reduction process included the following steps: zero-read subtraction, reference pixels correction, non-linearity correction, dark current subtraction, gain conversion, sky background subtraction, calibration, flat-field correction, and corrections for bad pixels and cosmic rays. For a detailed description of these steps, we refer the reader to the original Iraclis paper \citep{tsiaras_hd209}.

From the reduced spatially scanned spectroscopic images, the white (from 1.1-1.7 $\mu$m) and spectral light curves were subsequently extracted. The spectral light curves bands were selected such that the SNR is approximately uniform across the planetary spectrum. We then discarded the first orbit of each visit as they present stronger wavelength-dependent ramps, and the first exposure after each buffer dump as these contain significantly lower counts than subsequent exposures.

\begin{figure*}
    \centering
    \includegraphics[width=\columnwidth]{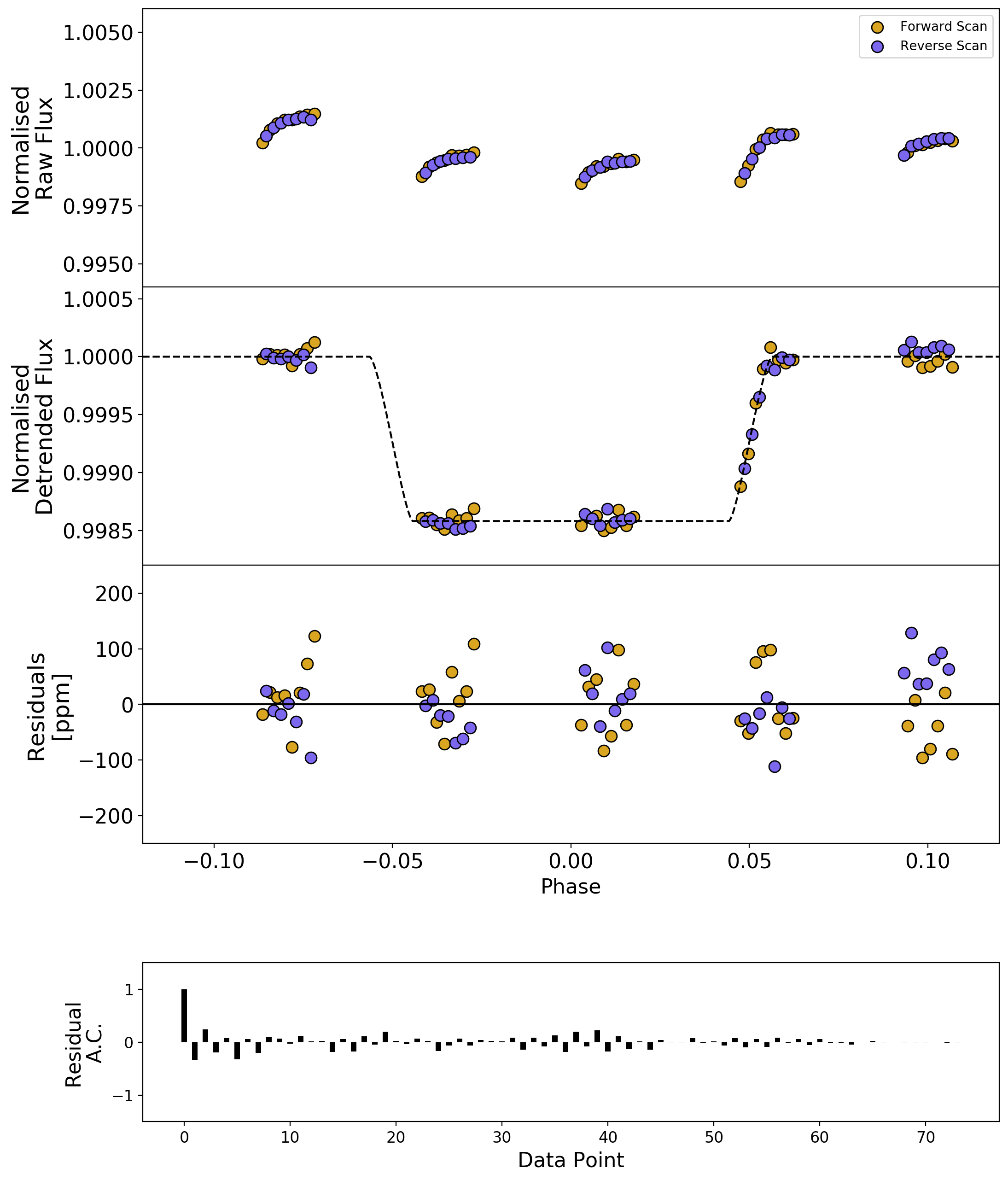}
    \includegraphics[width=\columnwidth]{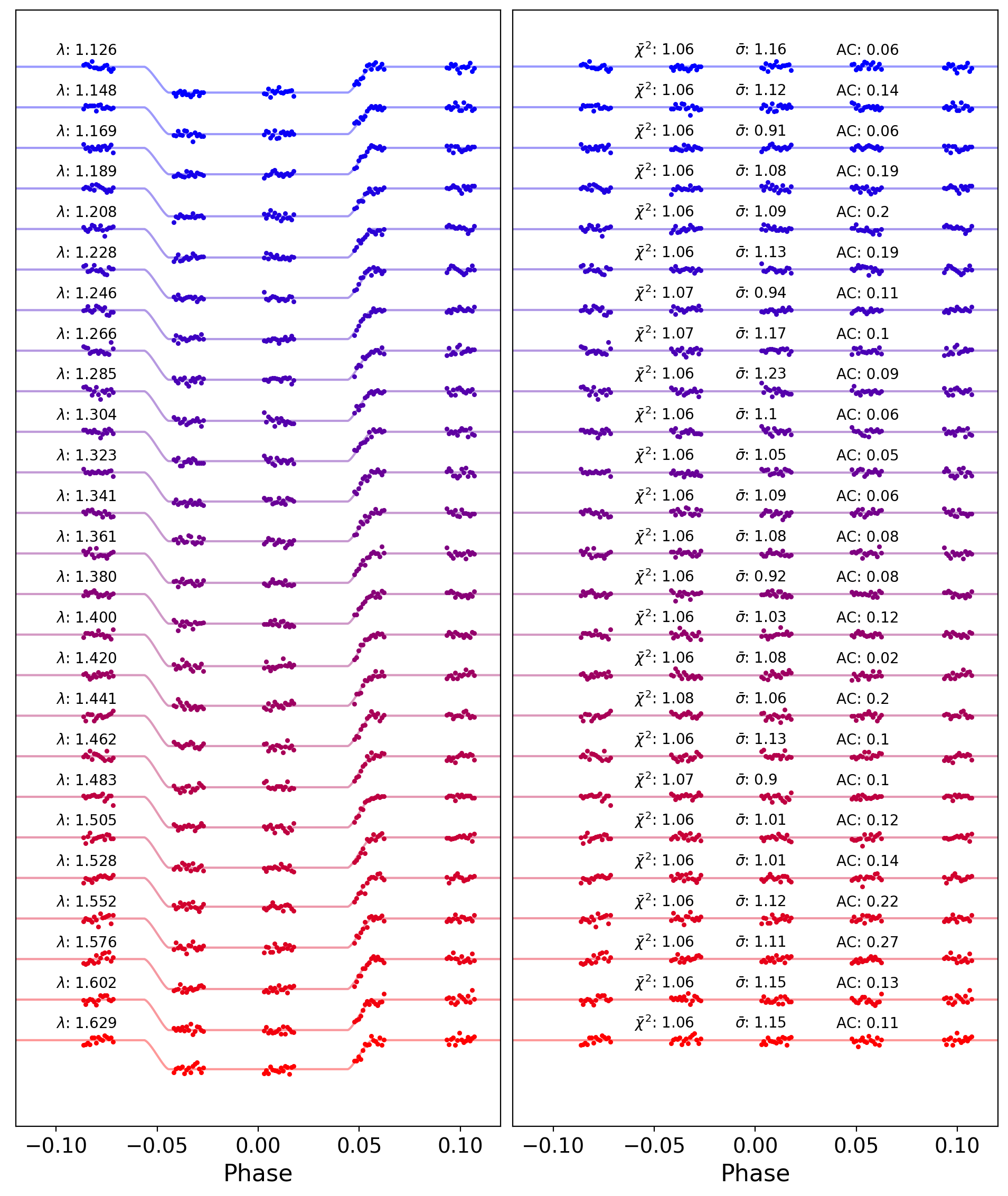}
    \includegraphics[width=\columnwidth]{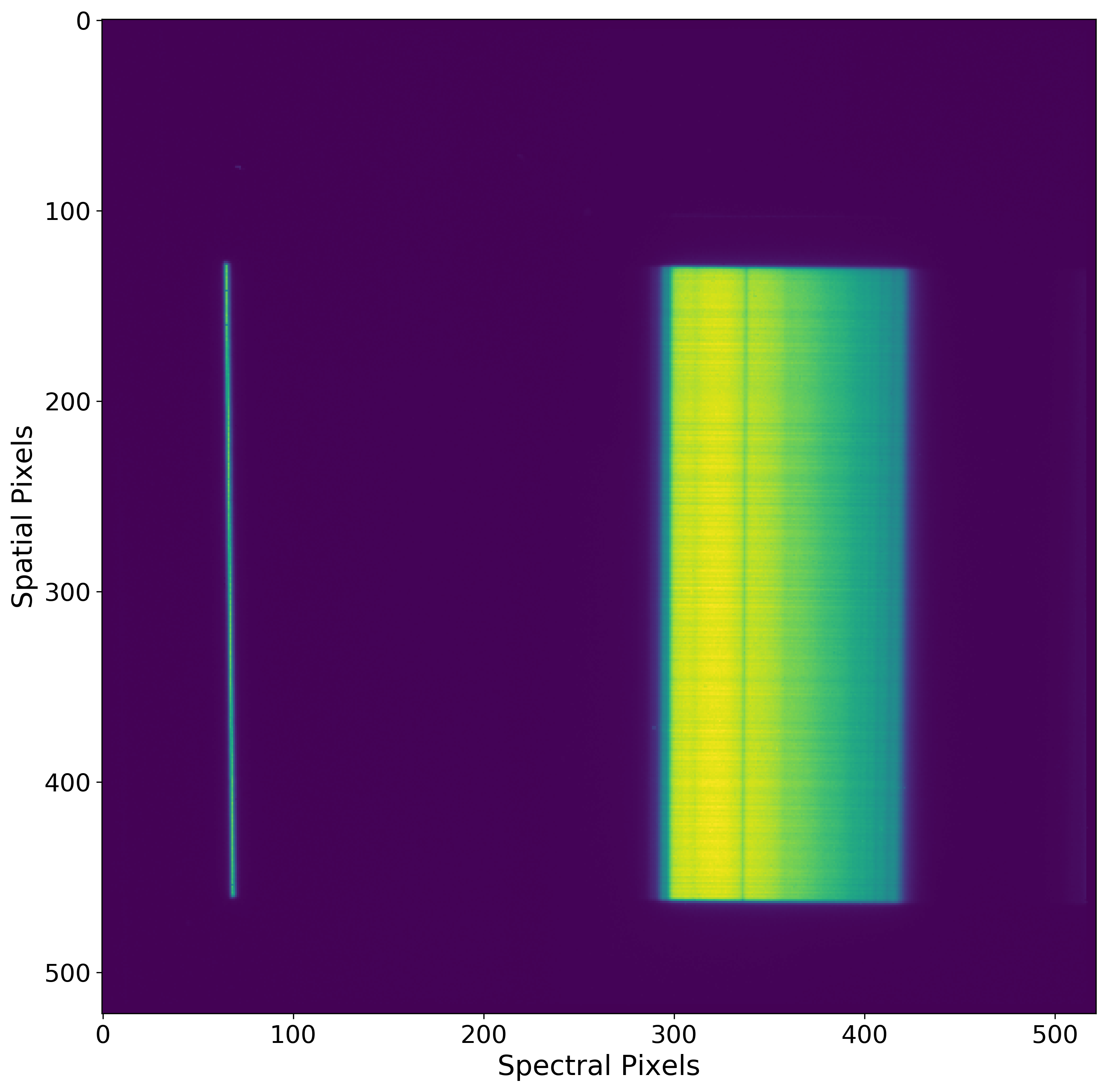}
    \includegraphics[width=\columnwidth]{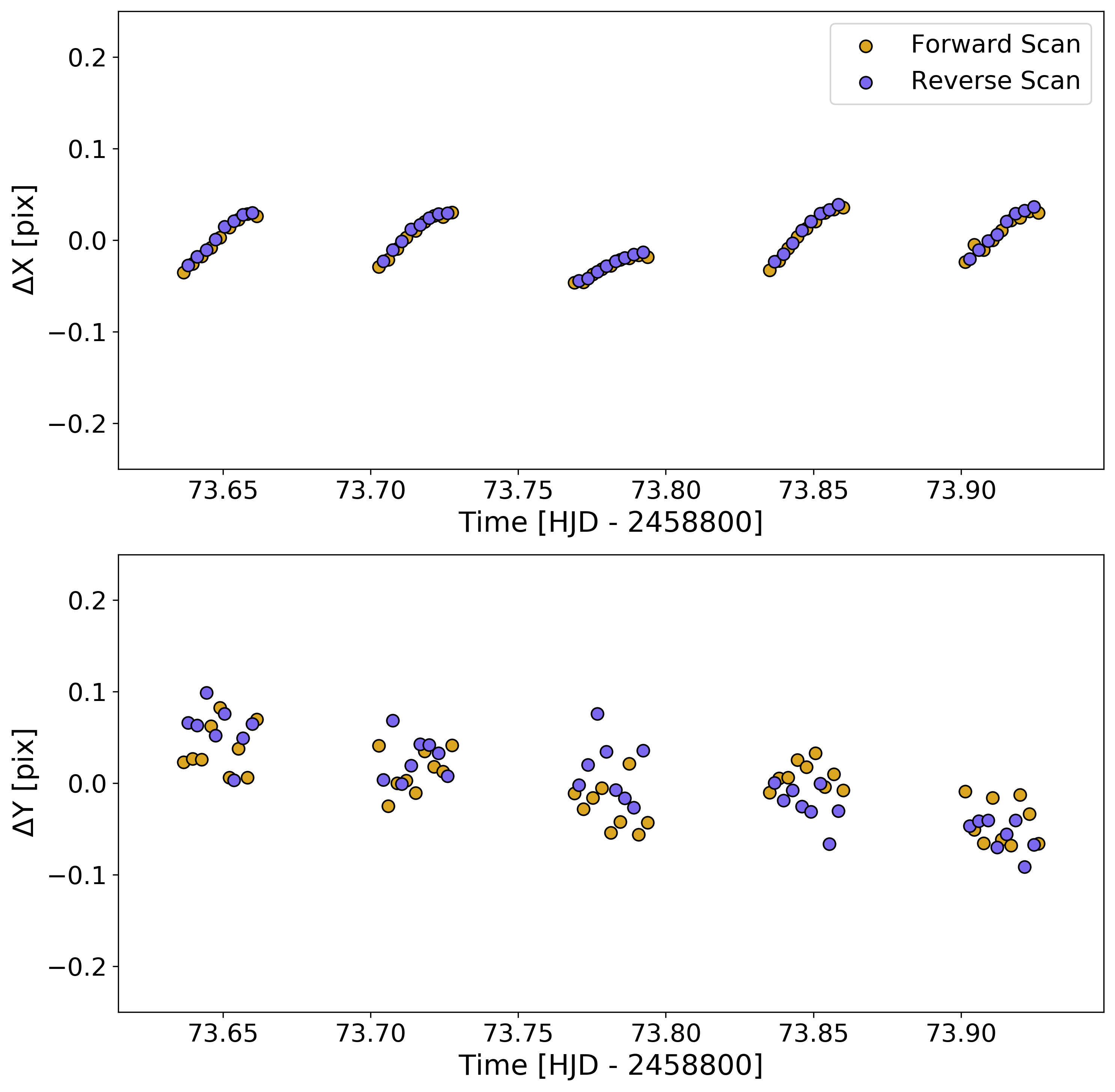}
    \caption{\textbf{Top left:} White light curve for the emission observations of KELT-9\,b. First panel: raw light curve, after normalisation. Second panel: light curve, divided by the best fit model for the systematics. Third panel: residuals for best-fit model. Fourth panel: auto-correlation function of the residuals. \textbf{Top right:} Spectral light curves fitted with Iraclis for the transmission spectra where, for clarity, an offset has been applied. Left panel: the detrended spectral light curves with best-fit model plotted. Right panel: residuals from the fitting with values for the Chi-squared ($\chi^2$), the standard deviation of the residuals with respect to the photon noise ($\bar{\sigma}$) and the auto-correlation (AC). \textbf{Bottom left:} Example detector image showing the large spatial scan. Both the first order, from which the spectrum is acquired, and the zeroth order can be seen. \textbf{Bottom Right:} Shifts in the X and Y location of each spectrum obtained.}
    \label{fig:light_curves}
\end{figure*}

We fitted the light curves using the PyLightcurve package\footnote{\url{https://github.com/ucl-exoplanets/pylightcurve}} which utilises the MCMC code ecmee \citep{emcee} and, for the fitting of the white light curve, the only free parameters were the mid-eclipse time and planet-to-star flux ratio. The other planet and the stellar parameters were fixed to the values from \citet{wong_k9} (a/R$_{*}$ = 3.191, i = 87.6$^{\circ}$, T$_{*}$ = 10170\,K). It is common in HST WFC3 data to have additional scatter that cannot be explained by the ramp model. For this reason, we scaled up the uncertainties in the individual data points, for their median to match the standard deviation of the residuals, and repeated the fitting, 

Subsequently we fitted the spectral light curves, with the mid-eclipse time fixed to that from the white fit and thus the eclipse depth was the only free parameter. The extracted eclipse spectrum is given in Table \ref{tab:spec} while the white and spectral light curve fits are given in Figures \ref{fig:light_curves}. Also shown are an example detector image and the shifts in the position of the spectrum on the detector over the course of the observation. Shifts in the positioning of the spectrum on the detector can cause significant systematics if uncorrected and \citet{stevenson_shifts} suggest that drifts which are smaller than 0.11 pixels (0.15 mas) in the spectral plane are optimal for exoplanet spectroscopy. The shifts seen here between observations are not significant and far smaller than those seen for 55\,Cancri\,e \citep{tsiaras_55cnce}.


\section{Atmospheric Retrieval}

Once the spectrum is obtained, we extracted the information content by performing a retrieval analysis using the open-source framework TauREx\,3 \citep{2019_al-refaie_taurex3}. The planet is simulated assuming a plane parallel atmosphere of 100 layers, spanning the pressure range from 10$^6$ to 10$^{-5}$ Pa in log space, assuming hydro-static equilibrium. Due to atmospheric outflows, the hydro-static equilibrium assumption might be invalid, especially at low densities. In this paper, due to the relatively low information content obtained spectra which limits the regions probed to high pressures, we do not consider more complicated modelling of those phenomena. In the retrievals, the bulk planet parameters (mass and radius) are fixed to the literature values (R$_{\rm p}$ = 1.89 R$_{\rm J}$, M$_{\rm p}$ = 2.88 M$_{\rm J}$, \citet{gaudi_k9}) due to the availability of more accurate constraints from transit and radial velocity techniques. We simulate the stellar spectrum from the literature parameters and the PHOENIX model \citep{Husser_2013}. In particular, for stars like KELT-9, a strong 1.28$\mu$m H I line is expected \citep{Malkan_2002}. Three free chemistry retrievals are performed: \\

$\bullet$ {\bf HST only:} In this scenario, only the newly reduced HST spectrum is considered. \\

$\bullet$ {\bf HST+Spitzer:} On top of the HST spectrum, we attempted to include the Spitzer data point from \citet{mansfield_k9} (E$_{\rm D}$ = 0.3131 $\pm$ 0.0062 \%). Previous studies highlighted the danger of such approach as there is no guarantee that these datasets are compatible \citep[e.g.][]{yip_lc, yip_wasp96, pluriel_aresIII, changeat_kelt11}. \\

$\bullet$ {\bf TESS+HST:} We also run a scenario including the TESS data from \citet{wong_k9} (E$_{\rm D}$ = 0.0065 $\pm$ 0.0015 \%) is added to the retrieval. Of course, this is subject to the same caveats as the HST+Spitzer case.\\

$\bullet$ {\bf TESS+HST+Spitzer:} For completeness, we also run a full scenario, which includes all the datasets. \\

In those retrievals, the abundances for H$_2$O \citep{polyansky_h2o}, TiO \citep{McKemmish_TiO_new}, VO \citep{mckemmish_vo}, CH$_4$ \citep{exomol_ch4}, CO \citep{li_co_2015}, FeH \citep{dulick_FeH, wende_FeH} and e$^-$ \citep{john_1988_h-} are considered constant with altitude and retrieved with large volume mixing priors between 10$^{-12}$ and 1. While more complex chemical profiles are potentially occurring in the atmosphere of KELT-9\,b, the retrieval of more complex parametric chemical profiles leads to large degeneracies in the case of HST \citep{Changeat_2019_2l}. In our retrievals, e$^-$ is used as a proxy for the H$^-$ opacity, following the implementation described in \cite{Edwards_2020_ares}. This description allows to calculate the bound-free and the free-free absorptions of H$^-$ when assuming the Saha equation, but also removes the free chemistry approach for this particular species. Since two degenerate free parameters are left (e$^-$ and H), we fixed the abundance of H using a 2-layer profile similar to what is done in \cite{Edwards_2020_ares}. We set the inflexion point of this profile at 1 bar, the mixing ratio for the surface at 10$^{-6}$ and the mixing ratio for the top of the atmosphere at 0.9. This profile is chosen to be consistent with the equilibrium predictions in Figure \ref{fig:eq_profiles} and leaves only the $e^-$ abundance as a free parameter. It is found that changing the assumption on the neutral H profile does not impact the main conclusions and only shifts the retrieved abundances slightly. The rest of the atmosphere is filled with hydrogen and helium assuming a solar composition (He/H = 0.17).


The thermal profiles are recovered using a free heuristic approach interpolating between 3 freely moving temperature-pressure points (NPoint profile). On top of the molecular absorbers, we include opacities from Collision Induced Absorption of the H$_2$-H$_2$ and H$_2$-He pairs as well as Rayleigh Scattering. The free parameters are explored using the Bayesian algorithm MultiNest \citep{Feroz_multinest,buchner_multinest} with an evidence tolerance of 0.5 and 1000 live points. For a more detailed description of the TauREx\,3 retrieval framework and the setup, see \cite{2019_al-refaie_taurex3}.

\begin{table}[]
    \centering
    \begin{tabular}{cccc}\hline\hline
    Wavelength [$\mu$m] & Depth [\%] & Error [\%] & Bandwidth [$\mu$m] \\ \hline
    1.12625 & 0.1275 & 0.0035 & 0.0219 \\
    1.14775 & 0.1324 & 0.0033 & 0.0211 \\
    1.16860 & 0.1304 & 0.0027 & 0.0206 \\
    1.18880 & 0.1377 & 0.0034 & 0.0198 \\
    1.20835 & 0.1408 & 0.0035 & 0.0193 \\
    1.22750 & 0.1401 & 0.0036 & 0.0190 \\
    1.24645 & 0.1431 & 0.0030 & 0.0189 \\
    1.26550 & 0.1436 & 0.0038 & 0.0192 \\
    1.28475 & 0.1463 & 0.0041 & 0.0193 \\
    1.30380 & 0.1449 & 0.0035 & 0.0188 \\
    1.32260 & 0.1440 & 0.0036 & 0.0188 \\
    1.34145 & 0.1400 & 0.0036 & 0.0189 \\
    1.36050 & 0.1451 & 0.0036 & 0.0192 \\
    1.38005 & 0.1463 & 0.0031 & 0.0199 \\
    1.40000 & 0.1560 & 0.0035 & 0.0200 \\
    1.42015 & 0.1514 & 0.0038 & 0.0203 \\
    1.44060 & 0.1500 & 0.0036 & 0.0206 \\
    1.46150 & 0.1532 & 0.0040 & 0.0212 \\
    1.48310 & 0.1510 & 0.0032 & 0.0220 \\
    1.50530 & 0.1498 & 0.0038 & 0.0224 \\
    1.52800 & 0.1421 & 0.0038 & 0.0230 \\
    1.55155 & 0.1438 & 0.0041 & 0.0241 \\
    1.57625 & 0.1481 & 0.0041 & 0.0253 \\
    1.60210 & 0.1505 & 0.0044 & 0.0264 \\
    1.62945 & 0.1434 & 0.0043 & 0.0283 \\\hline\hline
    \end{tabular}
    \caption{Extracted eclipse spectrum of KELT-9\,b}
    \label{tab:spec}
\end{table}

\section{Results}

\begin{figure*}
    \centering
    \includegraphics[width=\textwidth]{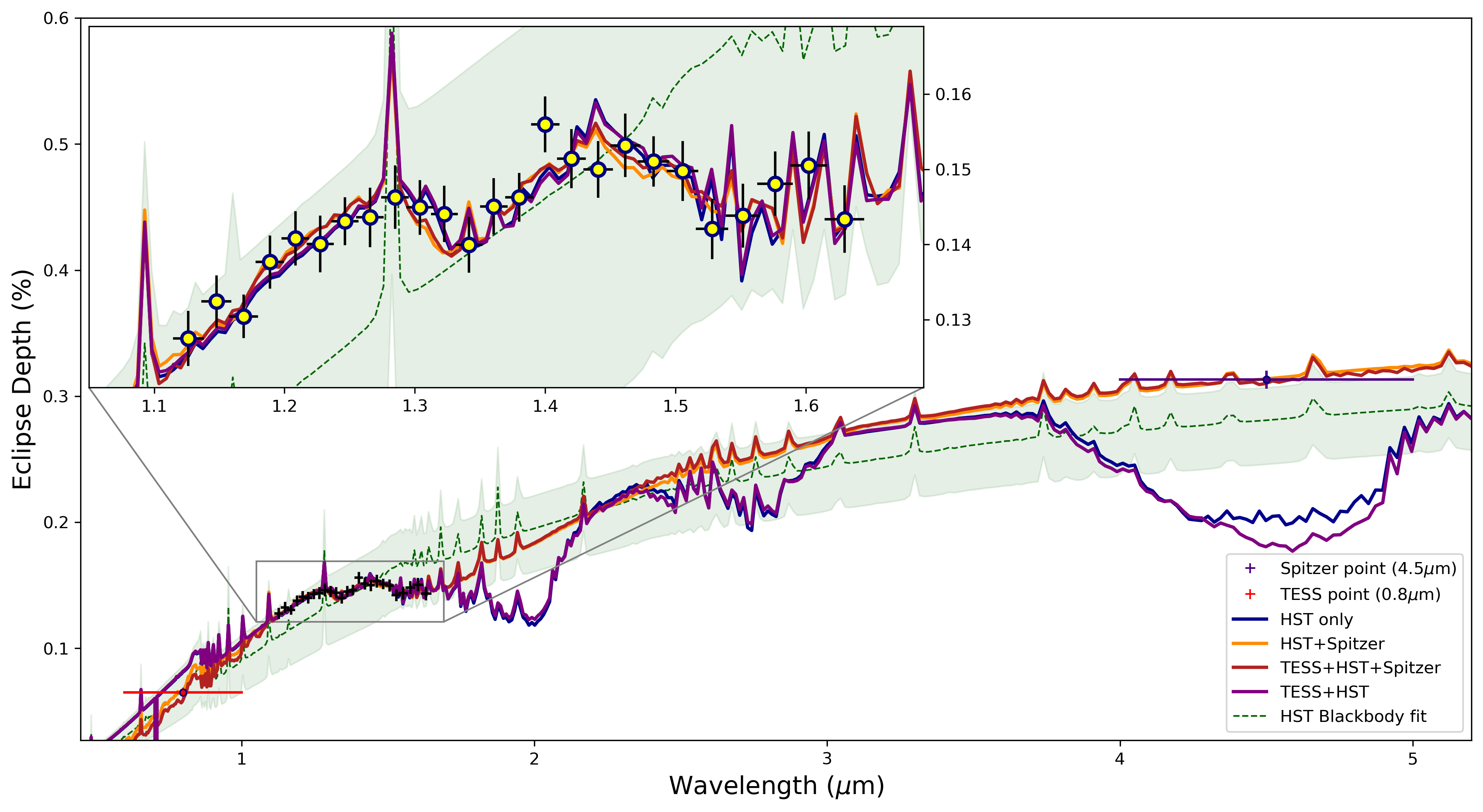}
    \caption{Extracted HST spectrum from our reduction with Iraclis (black), Spitzer photometric point (green) from \cite{mansfield_k9}, TESS photometric point (red) and the best-fit spectra from our three retrieval runs. We also provide in dashed green the best fit planet blackbody of the HST spectrum (T = 4504 K) as well as the region between the blackbody spectra at 4100K and 4800K in shaded green, illustrating the large feature sizes and differences between the HST, TESS and Spitzer datasets. Note that the H I line at 1.28 $\mu$m, as well as the other lines separated in the green blackbody fit, are from the PHOENIX model of the star \citep{Husser_2013}.}
    \label{fig:spectrum}
\end{figure*}

The resulting best fit spectra from our Iraclis reduction, from which we achieved an average precision of 36 ppm across 26 spectral channels, and the TauREx\,3 retrievals are provided in Figure \ref{fig:spectrum}. 

\begin{figure*}
    \centering
    \includegraphics[width=1\textwidth]{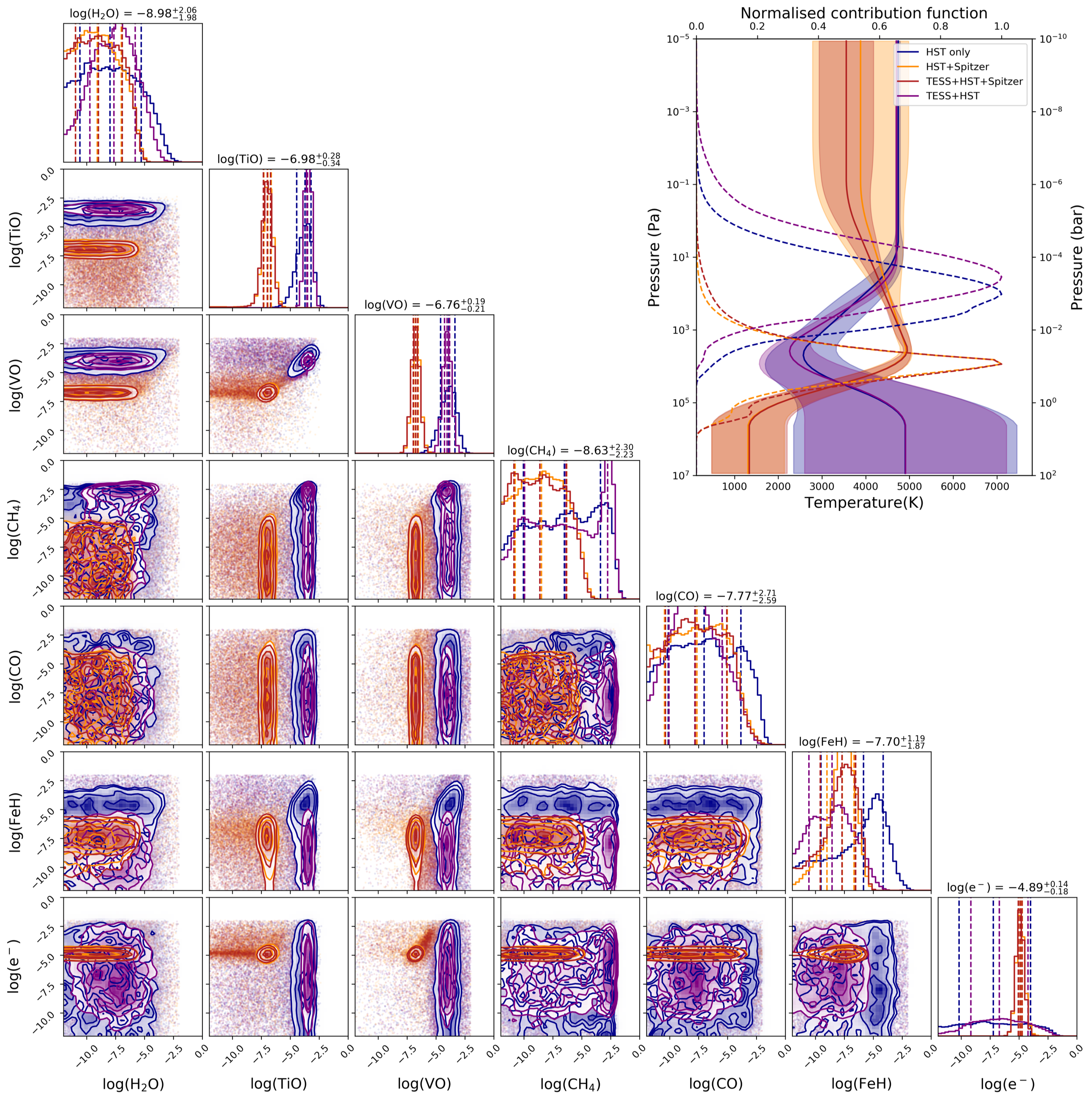}
    \caption{Retrieved posteriors and temperature-pressure profiles (top right) from our retrieval analyses. Blue: HST; orange: HST+Spitzer; red:TESS+HST+Spitzer; purple: TESS+HST. The contribution functions are also shown, highlighting two different classes of models, depending if Spitzer is included. Near-optical absorbers (TiO, VO, FeH and H-) are needed to explain the observed spectrum. When the TESS point is added, the retrieval results are almost identical. The values listed are from the TESS+HST+Spitzer retrieval.}
    \label{fig:posteriors}
\end{figure*}

In all three of our free retrievals, water vapour is constrained to abundances below 10$^{-5}$. Such a result is expected for planets of high temperatures like KELT-9\,b, where the majority of water is expected to be thermally dissociated \citep[e.g.][]{Parmentier_2018_w121photodiss}. Previous studies of this planet predicted that the day-side temperatures of KELT-9\,b should lead to enhanced thermal dissociation where molecules are not expected to survive. Despite this, the KELT-9\,b eclipse spectrum is consistent with the presence of emission features in the HST wavelength range and is poorly fitted by a simple planet blackbody (see green models in Figure \ref{fig:spectrum}). The observed features in the HST range are attributed to a mix of species with near optical absorbing properties: TiO, VO, FeH and H- (see Figure \ref{fig:posteriors}). When only the HST data is considered, the retrieved abundances for TiO and VO are high (respectively log(TiO)= $-3.7^{+0.5}_{-0.7}$ and log(VO)= $-3.9^{+0.5}_{-0.7}$), while FeH and H- are not detected. When the Spitzer photometric measurement at 4.5$\mu$m is included, the model converges to a more physical solution containing large abundances of e- (log(e-) = $-4.9^{+0.2}_{-0.2}$), a proxy for the H- absorption coming from the H$_2$ thermal dissociation \citep{Edwards_2020_ares}. On top of this, TiO is found at an abundance of log(TiO) = $-6.9^{+0.3}_{-0.4}$, VO at an abundance of log(VO) = $-6.7^{+0.2}_{-0.2}$ and FeH at log(FeH) = $-7.7^{+1.3}_{-1.3}$.

While the presence of molecular species remains surprising at those temperatures \citep{Woitke_2018}, this provides a much more convincing picture than the HST only retrieval. Besides, the observed HST spectrum contains large absorption features that cannot be fit with H- only and many atomic metals were already detected in the atmosphere of this planet from the cross-correlation technique \citep{hoeij_k9}. We highlight, however, that we can't guarantee the compatibility of the HST and Spitzer datasets, and that absolute offsets between those two datasets could lead to large biases in the retrieved abundances \citep[e.g.][]{yip_wasp96,pluriel_aresIII}. Nevertheless, we find that adding TESS to the retrieval, does not change much the retrieval results (see posterior distribution and temperature structure in Figure \ref{fig:posteriors}), which might provide indications that the datasets from TESS, HST and Spitzer are compatible for this planet. The presence of these molecular features in KELT-9\,b might suggest that disequilibrium processes and/or dynamical mixing with regions that are much cooler (e.g. the night-side \citet{wong_k9}) are allowing TiO, VO and FeH to remain stable in this atmosphere and be observed in this eclipse spectrum. For the other molecules included in the model (CO and CH$_4$), we do not recover any particular constraints. 

We quantified the statistical significance of the observed features in HST by comparing the logarithmic Bayesian evidence, log(E), of all our models with that of a baseline blackbody fit (log(E)$_{base}$ = 22.4 for HST alone, log(E)$_{base}$ = 23.8 for HST+Spitzer and log(E)$_{base}$ = 31.5 for TESS+HST+Spitzer). The blackbody temperatures associated with these baseline models are 4504 K, 4510 K and 4503 K respectively. In the free HST retrieval, the difference of log evidence is $\Delta$log(E) = 185.5. When the Spitzer point is accounted for, the difference is $\Delta$log(E) = 190.3 and when TESS is added, $\Delta$log(E) = 190.9. These strongly suggests that the observed features are not arising from statistical fluctuations \citep{Kass1995bayes} and that the presence of near-optical absorbers (TiO, VO, FeH and H-) is required to explain the spectrum of KELT-9\,b.

For the temperature structure, all the performed retrievals are consistent with the presence of a thermal inversion (see Figure \ref{fig:posteriors}). The thermal structure in the HST+Spitzer and the TESS+HST+Spitzer runs are almost identical, with an inversion relatively deep in the atmosphere (10$^5$ Pa). While cooler, the general temperature-pressure structure is consistent with the findings in \cite{fossati_2020_datadriven} and could allow for molecular species to survive in the deeper layers. The temperature reaches a maximum of about 5000 K. In the HST only case, we again find a thermal inversion although the temperature structure differs significantly with the inversion occurring far higher in the atmosphere (10$^3$ Pa). In Figure \ref{fig:posteriors} the contribution functions are also plotted, highlighting two classes of models where different regions are probed depending whether Spitzer is included or not. In either case, the retrievals explore a region where a thermal inversion occur, with the molecules being seen in emission. The inclusion of TESS only shifts the contribution up in the TESS+HST case, which is expected due the optical wavelengths probing much higher in the atmosphere.

The results for the TESS+HST run are also shown in the figures, but we find similar results to the HST only run in terms of molecular detection and thermal profile, indicating that the additional information contained in the TESS channel for this case is limited. We also investigated the significance of each species, by performing retrieval analyses on the TESS+HST+Spitzer case, removing the detected species one by one. The $\Delta$log(E) were 186.5, 176.1, 185.2, 181.2 for the runs without TiO, VO, FeH and H-, respectively. Given that the Bayesian evidence is significantly lower than the $\Delta$log(E) = 190.9 for the full model, the best statistical fit is the one containing all four species. However, we caution that this result does not prove that other metal oxide/hydrides and other absorbing species are not present in this atmosphere. In fact, we noticed during the removal process of the detected molecules, that the abundance of the remaining ones had the tendency to increase by a few orders of magnitude to compensate for the removed absorption, thus indicating that if other absorbers are indeed present, our retrieved abundances could be overestimated. Model dependent behaviour such as this one were thoroughly investigated in \citep{changeat_kelt11}. 


\section{Discussion}

The retrievals performed here explain the modular shape of the KELT-9\,b eclipse spectrum with a mix of near-optical absorbers. The prevalence of these molecules is higher than expected but we note that accurate abundances are difficult to extract from the low resolution, low wavelength coverage of HST, or even from combined datasets. In addition to this, the extreme nature of KELT-9\,b and its host star imply that conclusions must be cautiously drawn from analyses of these datasets.

A number of previous studies have found evidence for TiO, VO and H- using low resolution space-based emission spectroscopy from HST and Spitzer \citep[e.g.][]{Haynes_Wasp33b_spectrum_em,mansfield_hatp7,Evans_wasp121_e2,Edwards_2020_ares}. However, given the narrow spectral coverage and low resolution of the datasets, distinguishing between different opacity sources is difficult. Other hydrides and oxides that are not considered in this work (e.g. Aluminium Oxide \citep{von_essen_w33,chubb_w43}), may be contributing to the near-infrared features. The ones chosen here are the most likely from chemical models of ultra-hot Jupiters \citep{Parmentier_2018_w121photodiss,Lothringer_2018, kitzmann_k9} while other abundant species (e.g. SiO) do not have significant features in the WFC3 bandpass. Atomic and ionic species such as Fe, Fe$^+$ or Ti$^+$ were previously found in KELT-9\,b but since their electronic transitions occur in a much lower wavelength region than considered here ($< 0.8\mu$m), they were not included in our retrieval models \citep{Sharp_2007, Heiter_2015,Madhusudhan_2019, Ralchenko_2020}. Nevertheless, the possibility exists that the abundances of TiO, VO and FeH could be over estimated to account for additional absorption by missing species. Further data, covering shorter wavelengths, would help to investigate this scenario.

In addition to this, the accuracy of retrievals are directly linked to the quality and completeness of the available linelists. While many of the ExoMol linelists are suitably broad in pressure and temperature coverage for the majority of the currently-known exoplanet population, KELT-9\,b lies far outside the norms. The linelists of TiO \citep{McKemmish_TiO_new}, VO \citep{mckemmish_vo} and FeH \citep{dulick_FeH, wende_FeH} are computed for temperatures up to 3500 K. While we used the latest versions from \citep{chubb2020exomolop}, our retrievals push to temperatures exceeding 3500 K, forcing us to fix the opacities to the highest available temperatures, meaning that molecular lines might be missing. This could impact the accuracy of the retrievals presented here, potentially contributing in the higher than expected retrieved abundances, even though we are not probing individual lines. Future work by groups such ExoMol, HITEMP and HITRAN would valuable for the study of such hot planets.

For planets such as KELT-9\,b, the irradiation difference between the day and night-side is expected to induce a large day-night temperature contrasts: the difference is expected to be around 2000 K \citep{wong_k9, mansfield_k9}. These can lead to 3-dimensional biases \citep[e.g.][]{Taylor_2020, Feng_2020, Changeat_2020} due to the in-homogeneous day-side emission. In our retrieval, we employ a 1-dimensional description of the atmosphere, which might not well represent the actual planet. Additionally, in such a complex atmosphere, chemical species are not expected to be constant with altitude \citep[e.g.][]{Lothringer_2018}. Previous work from \cite{Changeat_2019_2l} highlighted that the small wavelength coverage of HST does not allow us to extract more complex chemical profiles unless additional assumptions such as equilibrium chemistry are adopted. Assuming such model to extract information content from exoplanet spectra is dangerous as it would only provide a model dependent solution, driven by the model assumptions.

Nevertheless, we also attempted an equilibrium chemistry retrieval using the same chemical scheme for the TESS+HST+Spitzer case. The results suggested that equilibrium chemistry might not a valid assumption for this planet: $\Delta$log(E) = 172.2 for the equilibrium run against $\Delta$log(E) = 189.8 for the free run. In addition to this, the equilibrium chemistry assumption lead to highly nonphysical retrieved parameters (log(Z) = 2.0$^{+0.1}_{-0.1}$, log(Ti/O) = 1.7$^{+0.1}_{-0.1}$ and log(V/O) = 1.9$^{+0.1}_{-0.1}$).

\begin{figure}
    \centering
    \includegraphics[width=0.47\textwidth]{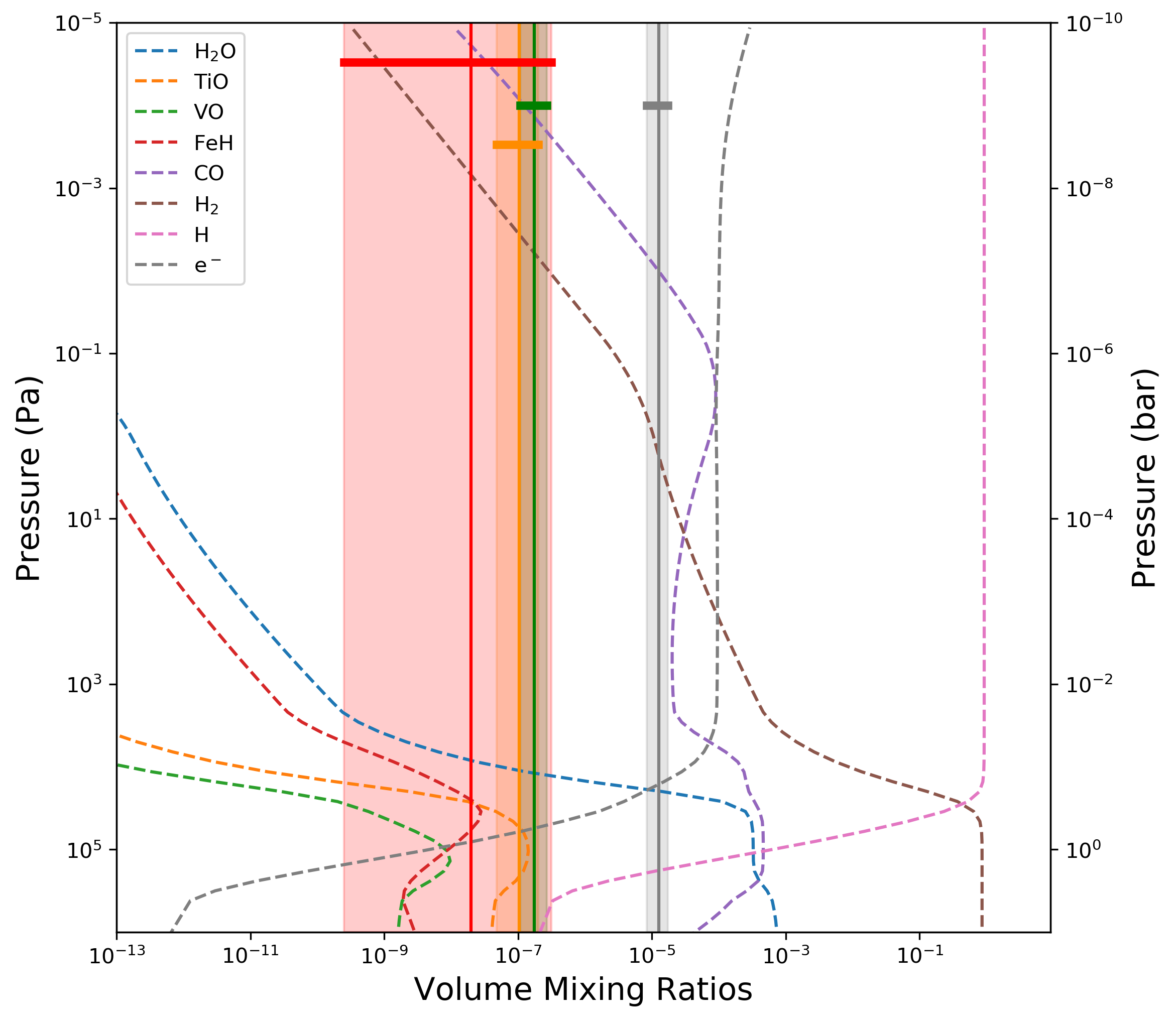}
    \caption{Chemical profiles for KELT-9\,b assuming equilibrium chemistry and the thermal profile from our TESS+HST+Spitzer fit (dotted lines) and retrieved chemistry from the same retrieval (shaded regions). The 1 sigma retrieved regions are also highlighted by the solid lines. The chemical scheme used is from \cite{Woitke_2018}.}
    \label{fig:eq_profiles}
\end{figure}

To compare the results from our free chemical retrievals with predictions from self-consistent models, we plot in Figure \ref{fig:eq_profiles} the chemical profiles obtained from assuming chemical equilibrium \citep[GGchem:][]{Woitke_2018} with solar abundances and the thermal profile obtained in the TESS+HST+Spitzer retrieval. 

We find that our retrieved H$_2$ dissociation in the TESS+HST+Spitzer case are consistent with the GGChem estimates, in the region probed by the observations. For TiO, VO and FeH, however, we consistently retrieved a higher than expected abundances. From the free chemical retrievals of the TESS+HST+Spitzer dataset, one finds the ratios Ti/H = $6.1\times10^{-8}$ and V/H = $1.0\times10^{-7}$, which are comparable with solar abundances if all the metals were in the form of those molecules. Given the ground based detections of atomic and ionic Ti \citep{hoeij_k9,hoeij_k9_2}, the high abundance of TiO and potentially VO found here suggests a super-solar metallicity or biases in our retrieved abundances. For Fe, we find that Fe/H = $1.2\times10^{-8}$, which is much lower than solar ratios. This ratio implies that the species must also be found in another form, which matches the findings in \cite{pino_k9} who detected atomic Fe, even though their assumed self-consistent temperature structure from \cite{Lothringer_2018} differs from our retrieved ones.

Disequilibrium processes, such as vertical or horizontal mixing could also play an important role in distributing metal oxide/hydrides on the day-side of KELT-9\,b \citep{kitzmann_k9, Komacek_2019, Drummond_2020}. To investigate those processes, one should compare the timescales of the creation and destruction reactions with the timescales of dynamical processes. However, little is known regarding TiO, VO and FeH kinetic reaction rate, limiting such studies, particularly for such an extreme object. If mixing processes are important, estimates on the dynamical timescales for such atmosphere suggest a lower limit on the destruction rate of the associated molecules between 10$^4$s and 10$^6$s \citep{Line_2011, kitzmann_k9}. However, due to the potential biases in the recovered molecular abundances, even with firm knowledge of these timescales, it would be difficult to ascertain whether these processes were occurring and were the cause of the increased abundances of metal oxides and hydrides.

Finally, the star KELT-9 is a fast rotating star with a non uniform structure that might not well be estimated from our PHOENIX models. While these effects would be more important in the transit case, a better understanding of the stellar properties might be needed to extract more precise information regarding KELT-9\,b.


Hence, the abundances of optical absorbers recovered here may well be overestimated. Nevertheless, the HST spectrum of KELT-9\,b clearly shows a strong deviation from a blackbody and, while the physical interpretation of these features via retrievals may have been affected by the reasons above, this in itself is unexpected suggesting the planet may be very different from our a priori assumptions.

\section{Conclusion}

We analysed the HST eclipse spectrum of KELT-9\,b, the hottest transiting planet discovered so far, using open source reduction and retrieval frameworks. While the extreme temperature of the planet's day-side was expected to induce thermal dissociation of the main molecular species and lead to an H- dominated spectrum with few broadband features, we find that the emission spectrum of KELT-9\,b contains rich features which imply the presence of molecular species. 

We explored 3 scenarios to ensure the consistency of our dataset across different combinations of instruments: HST only, TESS+HST, HST+Spitzer and TESS+HST+Spitzer. In all those scenarios, we find that the atmosphere must have a thermal inversion and that the observed features are well fitted by near optical absorbers (TiO and VO). Water vapour is not recovered in any of the investigated models, which is consistent with predictions of the molecule being dissociated. When Spitzer is added, the retrievals (with/without TESS) are almost identical and the presence of FeH and H- is also detected. In essence, these retrievals present a seemingly consistent picture, suggesting that the spectrum of KELT-9\,b cannot be fitted with a simple blackbody and that the presence of metal oxides and/or hydrides is required. 

The extreme nature of this planet mean that future investigations in terms of atmospheric modelling and line opacity calculations are required to overcome the limitations of this study. Nevertheless, the findings presented here contrast with the previous assumptions about the planet, bringing this extremely-hot Jupiter much closer to its ultra-hot Jupiter counterparts.

\vspace{3mm}

\textbf{Software:} Iraclis \citep{tsiaras_hd209}, TauREx3 \citep{2019_al-refaie_taurex3}, GGChem \citep{Woitke_2018}, Astropy \citep{astropy}, h5py \citep{hdf5_collette}, emcee \citep{emcee}, Matplotlib \citep{Hunter_matplotlib}, Multinest \citep{Feroz_multinest,buchner_multinest}, Pandas \citep{mckinney_pandas}, Numpy \citep{oliphant_numpy}, SciPy \citep{scipy}, corner \citep{corner}.

\vspace{3mm}
\textbf{Data:} This work is based upon observations with the NASA/ESA Hubble Space Telescope, obtained at the Space Telescope Science Institute (STScI) operated by AURA, Inc. The publicly available HST observations presented here were taken as part of proposal 15820, led by Lorenzo Pino \citep{Lorenzo_2019_proposals}. These were obtained from the Hubble Archive which is part of the Mikulski Archive for Space Telescopes.

\vspace{3mm}
\textbf{Acknowledgements:} This project has received funding from the European Research Council (ERC) under the European Union's Horizon 2020 research and innovation programme (grant agreement No 758892, ExoAI) and from the Science and Technology Funding Council (STFC) grant ST/S002634/1 and ST/T001836/1. We thank Giovanna Tinetti, Olivia Venot, Ahmed F. Al-Refaie, Sergey Yurchenko, Angelos Tsiaras and Jonathan Tennyson for their useful recommendations and discussions. We also wish to thank the reviewers, whom suggestions greatly improved this manuscript. Finally, we acknowledge the availability and support from the High Performance Computing platforms (HPC) DIRAC and OzSTAR, which provided the computing resources necessary to perform this work.

\bibliographystyle{aasjournal}
\bibliography{main}

\end{document}